\RequirePackage[2019-02-02]{latexrelease}
\documentclass{biometrika}

\usepackage{caption}
\usepackage[utf8]{inputenc}
\usepackage{amsmath, amssymb, graphicx, array, appendix, cleveref, enumitem, verbatim,subcaption,graphicx}
\usepackage[table,xcdraw]{xcolor}

\usepackage{times}
\usepackage{bm}
\usepackage{natbib}

\usepackage[plain,noend]{algorithm2e}

\makeatletter
\renewcommand{\algocf@captiontext}[2]{#1\algocf@typo. \AlCapFnt{}#2} 
\def\@algocf@capt@plain{top}
\renewcommand{\algocf@makecaption}[2]{%
  \addtolength{\hsize}{\algomargin}%
  \sbox\@tempboxa{\algocf@captiontext{#1}{#2}}%
  \ifdim\wd\@tempboxa >\hsize
    \hskip .5\algomargin%
    \parbox[t]{\hsize}{\algocf@captiontext{#1}{#2}}
  \else%
    \global\@minipagefalse%
    \hbox to\hsize{\box\@tempboxa}
  \fi%
  \addtolength{\hsize}{-\algomargin}%
}
\makeatother

\def\v{{\varepsilon}}

\def\hZ{\hat Z}
\def\hX{\hat X}
\def\hz{\hat z}

\def\hl{\hat \lambda}
\def\hL{\hat \Lambda}

\def\hV{\hat V}

\def\Eps{\mathcal{E}}

\def\cN{\mathcal{N}}
\def\Pone{P_{{1}}}
\def\v1{{1_n}}
\def\bM{\overline{M}}

\def\Cov{\mathrm{Cov}}

\def\P{\mathrm{P}}

\begin{document}

\jname{}


\markboth{Li et~al.}{High-dimensional Factor Analysis for Network-linked Data}

\title{High-dimensional Factor Analysis for Network-linked Data}

\author{Jinming Li, Gongjun Xu, and Ji Zhu}
\affil{Department of Statistics, University of Michigan\\  1085 South University Avenue, Ann Arbor, Michigan, 48109, U.S.A. \\
\email{lijinmin@umich.edu} \quad \email{gongjun@umich.edu} \quad \email{jizhu@umich.edu}}
\maketitle

\begin{abstract}
Factor analysis is a widely used statistical tool in many scientific disciplines, such as psychology, economics, and sociology. As observations linked by networks become increasingly common, incorporating network structures into factor analysis remains an open problem.
In this paper, we focus on high-dimensional factor analysis involving network-connected observations, and propose a generalized factor model with latent factors that account for both the network structure and the dependence structure among high-dimensional variables. These latent factors can be shared by the high-dimensional variables and the network, or exclusively applied to either of them. We develop a computationally efficient estimation procedure and establish asymptotic inferential theories. Notably, we show that by borrowing information from the network, the proposed estimator of the factor loading matrix achieves optimal asymptotic variance under much milder identifiability constraints than existing literature. Furthermore, we develop a hypothesis testing procedure to tackle the challenge of discerning the shared and individual latent factors' structure.
The finite sample performance of the proposed method is demonstrated through simulation studies and a real-world dataset involving a statistician co-authorship network. 
\end{abstract}




\section{Introduction}
Factor analysis is an important tool in multivariate data analysis \citep{anderson1956statistical, harman1976modern}, which characterizes the variability of correlated observed variables through a small number of common latent factors. It has been widely used in various fields, such as psychology \citep{ford1986application} and economics \citep{campbell1997econometrics}, where researchers are interested in exploring latent factors and interpreting their associations with observed variables. Driven by the emergence of large-scale data applications, recent research has focused on developing estimation procedures and inferential theories for factor models in high-dimensional settings \citep{bai2012statistical, bai2013principal, doz2012quasi, stock2002forecasting}.


Concurrently, advancements in technology have led to not only an increase in data dimensions but also in data variety. Applications involving both individual characteristics and relational information are becoming increasingly common. For instance, the National Longitudinal Study of Adolescent to Adult Health (AddHealth, \citealp{harris2009national}) documented demographic information of teenagers alongside their friendship networks. 
Although factor analysis has been used to analyze the relationship between questionnaire items and personalities in AddHealth data \citep{young2011measuring}, 
the authors did not utilize the friendship information. 
Other applications, such as online social media data from Facebook friendship networks \citep{snapnets}, include individual characteristics in conjunction with friendship information. For such network-linked data, incorporating network information into factor analysis remains an unresolved challenge.

In network analysis literature, latent space models have been used to integrate node variables when modeling network connections. { \color{black} However, these studies primarily focus on network modeling, treating node variables as supplementary information rather than the primary focus.} For instance, \cite{hoff2002latent}, \cite{ma2020universal}, and \cite{hoff2018additive} utilize both node variables and latent factors to model link probabilities. \cite{yang2013community} leverage node variables for community detection. { \color{black} \cite{zhang2020joint} propose a joint latent space model for network data and high-dimensional node variables. However, their model assumes identical latent factors shared between the two types of data, limiting its generalizability. More importantly, \cite{zhang2020joint} only study the average consistency property of the estimators, leaving uniform consistency, distributional properties, and statistical inference problems unexplored.
}

To tackle the challenge of modeling network-linked high-dimensional data, we introduce a generalized factor analysis model that simultaneously models high-dimensional node variables and network connections among observations. Specifically, 
we extend the classical factor analysis by not only considering latent factors associated with node variables but also introducing additional latent factors to explain the network's dependence structure.
Moreover, we provide flexibility by allowing latent factors to be either shared between the two types of data or specific to each type.
{ \color{black} The main contributions of this work are threefolds. 
First, from a modeling perspective, our proposed generalized factor model incorporates network information among the high-dimensional variables of interest, an aspect not yet explored in factor analysis literature. By integrating network information, we show that the generalized factor model requires much weaker identifiability conditions than existing results, thus necessitating fewer constraints for valid statistical inference. 
Second, from a statistical inference perspective, we propose a novel statistical hypothesis testing procedure to infer the shared and individual latent factors. We also establish the asymptotic distribution results for the latent factor and factor loading estimators, showing that the optimal asymptotic variance is achieved under weaker identifiability conditions compared to existing studies in factor analysis literature  \citep[e.g.,][]{bai2012statistical}. Numerical studies support our theory and further indicate that the proposed testing procedure achieves performance comparable to an oracle with known latent factors.
Lastly, from a statistical learning perspective, the model's flexibility, which allows latent factors to be shared between the high-dimensional variables and the network or to be exclusive to either, can lead to improved performance in downstream tasks,
which will be evidenced by our simulation studies and data analysis. 

} 


\section{The Generalized Factor Model}
\label{sec:the_joint_model_framework}

\subsection{Model Setup}
\begin{figure}[h]
    \centering
    \includegraphics[scale=0.2]{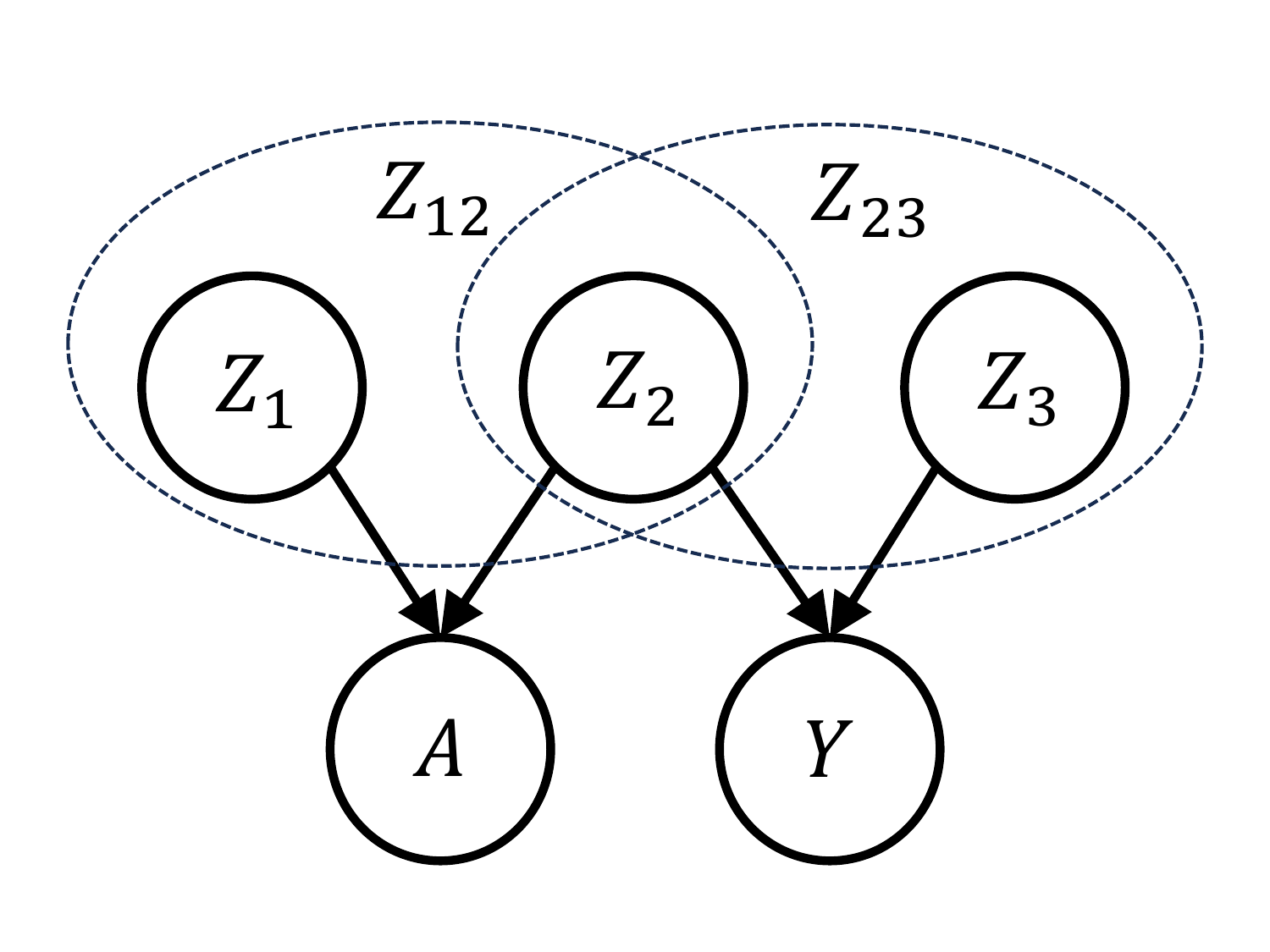}
    \caption{{\color{black} Illustration of observed data and latent factors.}}
    \label{fig:dependency_figure_LSM}
\end{figure}

Let $Y_{n \times p}$ represent the data matrix of $p$ high-dimensional variables for $n$ observations, where $y_i$ denotes the $i$-th row (expressed as a $p\times 1$ vector). Let $A_{n \times n}$ denote the binary adjacency matrix containing  network link information among these observations, where $A_{ij} = 1$ if nodes $i$ and $j$ are connected and $A_{ij} = 0$ otherwise. The latent factors for the two types of data are divided into three parts, denoted as  $Z_{n \times k} = \begin{pmatrix}Z_1 & Z_2 & Z_3 \end{pmatrix}$, with column dimensions $k_1, k_2, k_3$ respectively and $k = k_1 + k_2 + k_3$. {\color{black}\Cref{fig:dependency_figure_LSM} illustrates the dependency structure between the observed data and the latent factors:} $Z_1$ are latent factors related solely to network data, $Z_3$ are associated only with node variables, while $Z_2$ represent the overlapped factors shared by both types of data. We use $Z_{12} = \begin{pmatrix}Z_1 & Z_2 \end{pmatrix}$ and $Z_{23} = \begin{pmatrix}Z_2 & Z_3 \end{pmatrix}$ to denote the corresponding submatrices of $Z$. Following \cite{bai2012statistical}, we assume that rows of $Z$, denoted as $\{z_i\}_{i=1}^n$, are fixed constants. If $\{z_i\}_{i=1}^n$ were assumed to be a series of random variables, our analysis could be interpreted as conditioning on them. A more detailed discussion is provided in \Cref{sec:theoretical_analysis}.

We now introduce the proposed joint model for both types of data. The high-dimensional variables are assumed to follow a factor analysis model, and for each observation $i$, we have $$y_{i} = \mu + \Lambda z_{i, 23} + \epsilon_i,$$ where $\mu_{p \times 1}$ represents the mean vector, $\Lambda_{p \times (k_2 + k_3)}$ is the factor loading matrix, $z_{i,23}$ denotes the $i$-th row of $Z_{23}$ (expressed as a $(k_2+k_3)\times 1$ vector), and $\epsilon_i$'s are i.i.d. random noise following some unknown distribution with zero mean and diagonal covariance matrix $\Psi_{p \times p}$. For network links, we consider an inner product latent space model where $A_{ij}$'s follow independent Bernoulli distributions with probability $$P_{ij} = \alpha_i + \alpha_j + z_{i,12}^T z_{j,12},$$ such that $P_{ij} \in [0, 1]$. Under this model, two observations have a higher connecting probability if their latent positions exhibit a higher inner product similarity, or larger scalar parameters $\alpha_i$ and $\alpha_j$ which capture observation-specific degree heterogeneity. This network model closely relates to the Random Dot Product Graph model (RDPG) (\citealp{young2007random}), which unifies many popular latent space models in network literature, such as the Stochastic Block Model (SBM) (\citealp{holland1983stochastic}) and Degree Corrected Stochastic Block Model (DCSBM) (\citealp{karrer2011stochastic}). Additionally, the model can be viewed as an inner product latent space model proposed by \cite{ma2020universal} with an identity link function. More detailed discussion is available in the supplementary materials.

Interestingly, the proposed network model can also be interpreted as a factor model for binary relational data. In this interpretation, each observation has $n-1$ binary variables that correspond to its linkage with the remaining observations. Here, $Z_{12}$ serves the dual role of latent factors and factor loadings, while $\alpha_{n \times 1}$, analogous to $\mu_{p \times 1}$ in the factor model, accounts for the observation-wise mean structure. This interpretation aligns with the perspective presented by \cite{hoff2018additive}. 

Overall, the matrix form of the generalized factor model can be concisely represented as:
\begin{gather}
    \label{model:fa}
    Y = 1_{n} \mu^T + Z_{23}\Lambda^T + \mathcal{E}, ~\epsilon_{i} \overset{i.i.d.}{\sim} (0, \Psi),\\
    \label{model:net}
    A_{ij} \overset{ind.}{\sim} Bernoulli(P_{ij}) \mbox{ with } P = \alpha 1_{n}^T + 1_{n} \alpha^T + Z_{12}Z_{12}^T,
\end{gather}
 where $1_{n}$ denotes the $n\times 1$ all-ones vector and $\mathcal{E}_{n \times p}$ is the error matrix containing all $\epsilon_i$'s. 


\subsection{Identifiability Conditions}
\label{sec:identifiability}
Our goal is to estimate all model parameters $(\mu, \Lambda, Z, \Psi, \alpha)$. To do so, we first examine the identifiability conditions for \eqref{model:fa} and \eqref{model:net} separately and then discuss the general case involving shared factors. The main result reveals that by incorporating network information, we can reduce the number of identifiability conditions by $k_2 (k_2 + 1) / 2$.
We derive the identifiability conditions under the following setting.

\begin{definition}[Identifiability of the generalized factor model]
The model is identifiable if, for any two sets of model parameters $(\mu, \Lambda, Z, \Psi, \alpha)$ and $(\mu^*, \Lambda^*, Z^*, \Psi^*, \alpha^*)$ that satisfy the following equations:
\begin{gather}
    P = \alpha 1_{n}^T + 1_{n} \alpha^T + Z_{12} Z_{12}^T = \alpha^* 1_{n}^T + 1_{n} {\alpha^*}^T + Z_{12}^* {Z_{12}^*}^T, \label{idf:net}
\\
    \Cov(Y) = \Lambda M_{23} \Lambda^T + \Psi = \Lambda^* M_{23}^* {\Lambda^*}^T + \Psi^*, \label{idf:fa_ori}
\end{gather}
where $M_{23} = Z_{23}^TZ_{23} / n, M_{23}^* = {Z_{23}^*}^T Z_{23}^*/n$ are covariance matrices, it must be the case that $\mu = \mu^*, \Lambda = \Lambda^*, Z = Z^*, \Psi = \Psi^*, \alpha = \alpha^*$.
\label{def:identifiability}
\end{definition}

Considering \eqref{idf:net} and \eqref{idf:fa_ori} separately, we first need $Z$ to be centered to identify $\mu$ and $\alpha$. Then, $Z_{12}$ is identifiable up to any rotation matrix $O$, while $Z_{23}$ and $\Lambda$ are identifiable up to any invertible matrix $Q$, as we can see from the equations below:
\begin{gather*}
    Z_{12}Z_{12}^T = Z_{12}OO^TZ_{12}^T = (Z_{12}O)(Z_{12}O)^T,
    \label{eq:oth1}
 \\
 \Lambda M_{23} \Lambda^T = (\Lambda Q) (Q^{-1} M_{23} Q^{-T}) (\Lambda Q)^T = (\Lambda Q) (Q^{-1} Z_{23}^T Z_{23} Q^{-T}/n) (\Lambda Q)^T.
    \label{eq:oth2}
\end{gather*}
Existing literature in factor analysis has proposed several sets of identifiability conditions to fix $Q$, as summarized in \cite{bai2012statistical}. For convenience, we list these conditions (IC*1-5) in \Cref{tab:ic_bai}.

\begin{table}[h]
    \centering
    \begin{tabular}{|c|c|c|}
        \hline
            & $M_{Z}$ & $\Lambda$\\
        \hline
        IC*1 & No restriction & Contains submatrix $I_k$\\
        IC*2 & Diagonal with distinct elements & $\Lambda^T \Psi^{-1} \Lambda/p = I_k$\\
        IC*3 & Identity matrix & $\Lambda^T \Psi^{-1} \Lambda/p$ diagonal with distinct elements\\
        IC*4 & Diagonal matrix & Lower triangular with diagonal elements 1\\
        IC*5 & Identity matrix & Lower triangular \\
        \hline
    \end{tabular}
    \caption{Identifiability Conditions (Factor Model).}
    \label{tab:ic_bai}
\end{table}

First, consider the simple situation where all factors are shared, i.e., $Z = Z_{2} = Z_{12} = Z_{23}$. Intuitively, we can reduce the number of identifiability conditions for $Z_2$ and $\Lambda$ by introducing network information with the following observations: orthogonal matrices for the network model and the factor model are identical, i.e., $O = Q$, and $ZZ^T$ shares the same eigenvalues with $Z^TZ$. We formalize these observations in the following theorem.

\begin{theorem}
\label{thm:identifiability}
    If $Z = Z_2$, then by having $Z$ being centered and any set of conditions in \Cref{tab:ic} (IC1, IC2, or IC3), we can achieve strict identifiability as per \Cref{def:identifiability}.
\end{theorem}

\begin{table}[h]
    \centering
    \begin{tabular}{|c|c|c|}
        \hline
            & $M_{Z}$ & $\Lambda$\\
        \hline
        IC1 & No restriction & $\Lambda^T \Psi^{-1} \Lambda/p$  diagonal with distinct elements \\
        IC2 & Diagonal with distinct elements & No restriction \\
        IC3 & Distinct eigenvalues & Lower triangular \\
        \hline
    \end{tabular}
    \caption{Identifiability Conditions (Generalized Factor Model).}
    \label{tab:ic}
\end{table}

By comparing the identifiability conditions in \Cref{tab:ic_bai} and \Cref{tab:ic}, we can observe that for the shared factors $Z_2$, the number of conditions required is reduced by $k_2(k_2+1)/2$. This suggests that the generalized factor model necessitates fewer constraints for valid inference and is more accommodating for experimental design in confirmatory factor analysis. The incorporation of network information into the factor analysis model provides a significant advantage in this regard.

We further extend the result in \Cref{thm:identifiability} to the general case where $Z_1$ and $Z_3$ also exist. As anticipated, while the number of conditions can be reduced for shared factors $Z_2$, additional identifiability conditions are required for individual factors $Z_1$ and $Z_3$. Generally, one needs $k(k-1)/2$ conditions to resolve the rotation indeterminacy of $Z$, and $k_3(k_3 + 1)/2$ conditions to address the indeterminacy of $\Lambda_3$. {\color{black} Practically, the following sets of conditions are natural for interpretation and compatible with the two-step estimation procedure described later. Nevertheless, we acknowledge that these are not the only sets of conditions that achieve model identifiability.}

\begin{corollary}\label{cor:identifiability_of_general_joint_model}
In the general case where $Z_1, Z_2$ and $Z_3$ all exist, the generalized factor model is identifiable if one of the following sets of conditions is satisfied, each corresponds to a condition in \Cref{tab:ic}: (1) $Z$ is centered, $Z_{12}^TZ_3 = 0_{(k_1 + k_2) \times k_3}$, $\Lambda_2^T\Psi^{-1}\Lambda_2/p$ is diagonal with distinct elements, and $\Lambda_3$ contains submatrix $I_{k_3}$; (2) $Z$ is centered, $Z^TZ$ is diagonal with distinct elements, and $\Lambda_3^T\Psi^{-1}\Lambda_3/p = I_{k_3}$; (3) $Z$ is centered, $Z_{12}^TZ_3 = 0_{(k_1 + k_2) \times k_3}$, $Z_{12}^TZ_{12}$ has distinct eigenvalues, $Z_3^TZ_3$ is diagonal with distinct eigenvalues, $\Lambda_2$ is lower triangular, and $\Lambda_3^T\Psi^{-1}\Lambda_3/p = I_{k_3}$.
\end{corollary}

Finally, for all  the identifiability conditions in \Cref{tab:ic_bai}, \Cref{tab:ic} and \Cref{cor:identifiability_of_general_joint_model}, one needs to further specify column signs to achieve strict identifiability. A straight forward condition is to set the first non-zero entry of each column of $Z$ to be positive. 

\subsection{Estimation Approach}
\label{sec:estimating_model_parameters}

In this section, we introduce our estimation method, focusing on the situation where $k_1, k_2$ and $k_3$ are known, while addressing the issue of estimating them in \Cref{sec:hypothesis_testing_of_overlapped_latent_structure}. 
We propose a two-step estimation approach that is computationally efficient for large-scale data and offers desirable theoretical properties for the estimators. 
Specifically, the proposed two-step estimation approach proceeds as follows: (1) obtain the estimator $\hZ_{12}$ from network data; and (2) regresses high-dimensional variables on $\hZ_{12}$ to estimate factor loadings. We demonstrate that $\hZ_{12}$ is accurate enough to ensure that the factor structure related to $Z_3$ is well preserved within regression residuals and can then be estimated using a standard factor analysis method.

\smallskip 
\label{sec:estimate_network_model}

\textit{Step 1: Estimating the Network Model.} We utilize adjacency spectral embedding (ASE) (\citealp{athreya2017statistical}) to estimate the network model. Specifically, we use the $d$-dimensional ASE of $A$, denoted by $U_AS_A^{1/2}$, where $S_A^{1/2}$ is a diagonal matrix with entries being the square-roots of the $d$ largest singular values of $A$, arranged in descending order, and $U_A$ is the $n \times d$ matrix with corresponding orthonormal columns. Note that $U_AS_AU_A^T$ provides a low-rank approximation of $A$, while $A$ can be viewed as a random realization of $P$. Thus, centering on \eqref{model:net}, the equation $(I - \Pone)P (I - \Pone) = Z_{12}Z_{12}^T$  leads to a natural estimator of $Z_{12}$ as the centered ASE of $A$, $ \hZ_{12} = (I - \Pone)U_AS_A^{1/2}$, where $I$ is the identity matrix and $\Pone$ is the projection matrix induced by the vector $1_{n}$. Following \cite{athreya2017statistical}, the dimension $d = k_1 + k_2$ can be chosen with a scree-plot-based method \citep{zhu2006automatic}. To satisfy identifiability conditions, one can further perform rotation or sign-changing on $\hZ_{12}$. On the other hand, with the equation $\Pone P \Pone = \alpha 1_{n}^T + 1_{n} \alpha^T$, we can derive an estimate for node-specific parameters $\alpha$ as $\hat{\alpha} = n^{-1}\{I - P_{1}/2\}A1_{n}$.

\label{sec:estimate_factor_model}
\textit{Step 2: Estimating the Factor Model.} We regress $Y$ on $\hZ_{12}$ to estimate the factor model. Since $Z$ is centered by identifiability conditions, we have $\hat{\mu} = Y^T 1_{n} / n$. For notational simplicity, in the following discussion, we omit the estimation of $\mu$ by directly assuming $Y$ is centered. While $\hZ_{12}$ is estimated from the first step, it is still unknown whether a factor, or a column of $\hZ_{12}$, belongs to $Z_1$ or $Z_2$. We include $Z_1$ in the factor model and rewrite \eqref{model:fa} as
\begin{equation}
    Y = 1_{n} \mu^T + Z_1 0_{k_1 \times p} +  Z_{2}\Lambda_2^T + Z_{3}\Lambda_3^T + \mathcal{E}.
\label{eq:linear_regression}
\end{equation}
Thus, we can estimate $\Lambda_{12}$ by performing OLS regression of $Y$ with $\hZ_{12}$ as a plug-in estimator of $Z_{12}$, and the corresponding estimator $\hL_{12}$ is 
{\color{black} $\hL_{12}^T = (\hZ_{12}^T \hZ_{12})^{-1} \hZ_{12}^T Y$}.

Next, we estimate $Z_3, \Lambda_3$ and $\Psi$ from the residuals of the regression, while leaving  the issue of differentiating $Z_1$ and $Z_2$ for \Cref{sec:hypothesis_testing_of_overlapped_latent_structure}.
Denote $R = (I - \Pone - \P_{\hZ_{12}})Y$ as the residual matrix and $\P_{\hZ_{12}} = {\hZ_{12}}({\hZ_{12}}^T{\hZ_{12}})^{-1}{\hZ_{12}}^T$ as the projection matrix induced by $\hZ_{12}$. We propose the following estimation procedure: if $Z_3$ does not exist, i.e. $k_3 = 0$, we directly estimate $\Psi$ by $\hat\Psi = \Cov(R)$; otherwise, we can apply existing estimation methods of high-dimensional factor models on $R$ to obtain $\hZ_3, \hL_3$, and $\hat\Psi$. In particular, we implement the estimation procedure in \cite{bai2012statistical} that uses the EM algorithm to minimize the following quasi log-likelihood function:
\begin{equation*}
L(Z_3, \Lambda_3, \Psi)=p^{-1}\left\{\ln \left|\Sigma_{Y_3}\right| + \operatorname{tr}\left(M_{Y_3} \Sigma_{Y_3}^{-1}\right)\right\},
\label{eq:loss_in_bai}
\end{equation*}
where $\Sigma_{Y_3} = \Lambda_3 M_{3} \Lambda_3^T + \Psi$, $\left|\cdot\right|$ and $\operatorname{tr}$ are the determinant and trace operators, respectively, $M_{Y_3} = Y_3^TY_3/n$, and $M_{3} = Z_3^TZ_3/n$. The problem of detecting the existence of $Z_3$ will be discussed in \Cref{sec:hypothesis_testing_of_overlapped_latent_structure}. {\color{black}The overall estimation and inference procedure is summarized in the Algorithm \ref{algo:summary_procedure}.}

\section{Theoretical Analysis}
\label{sec:theoretical_analysis}

\subsection{Asymptotic Properties of Network Model Estimators}

In this section, we present the asymptotic properties of the estimators derived from the two-step estimation procedure.
{\color{black}
In the related literature \citep{zhang2020joint}, only the average consistency properties have been studied, which is insufficient for conducting statistical inference on the factors and loadings—an important yet underexplored question. In contrast, our work establishes not only the average consistency rates but further the uniform consistency, and more importantly the asymptotic distributions.
}
Similar to the setting in \Cref{sec:estimating_model_parameters}, we assume that $k_1, k_2$ and $k_3$ are known. Throughout the analysis, we also make the following assumptions:
\renewcommand{\theenumi}{\Roman{enumi}}
\begin{enumerate}
\label{item:assumptions}
\item The latent factors $Z$ are treated as a series of fixed constants, with the rank-$k$ asymptotic covariance $\bM_{Z} {=} \lim_{n \rightarrow \infty} Z^T Z/n$ existing.
\item The adjacency matrix $A$ is symmetric with all diagonal entries being $0$.
\item $P$ is positive-semidefinite, and $\max_{1\leq i \leq n} \sum_{j = 1}^n P_{ij} > (\log n)^{4+a}$ for some $a > 0$.
\item There exists a constant $C>0$ large enough such that $0 < \Psi_{jj} \leq C$, $\|\lambda_{j}\|_{2} \leq C$ hold for all $1 \leq j \leq p$, and $\max_{i,j}|(\bM_{Z})_{ij}|\leq C$, where $\lambda_j$ is the $j$-th row of $\Lambda$.
\item The joint model satisfies any set of identifiability conditions described in \Cref{cor:identifiability_of_general_joint_model}.
\end{enumerate}

 Assumption I, following  \cite{bai2012statistical}, assumes $Z$ as a sequence of constants, which is much weaker compared to imposing distributional assumptions on $Z$. While latent factors $Z$ are treated as fixed constants in this work, our approach can also be applied to cases where $Z$ are treated as random variables. In such cases, our theoretical results can be regarded as being derived by conditioning on $Z$. More specifically, $\hat{Z}$ will still be consistent at the same rate, but the asymptotic variance would be related to the distribution of $Z$ and would need further derivation. Similar identifiability conditions can also be proposed.
Assumptions II and III are regularity conditions following the spectral embedding literature \citep{athreya2017statistical}, with Assumption III ensuring that the largest expected node degree of the network is not too small. By assuming $P$ to be positive-semidefinite, model \eqref{model:net} is equivalent to the Random Dot Product Graph (RDPG) model, and many popular block models in network analysis can be seen as special cases of our model. A more detailed discussion is provided in the supplementary materials. Assumption IV contains regularity conditions such that the factor model is well-defined in the  high-dimensional setting \citep{bai2012statistical}. Assumption V ensures that all parameters are identifiable and, hence, estimable. {\color{black} In particular, with Assumption V the following developed statistical inference results will not have rotation indeterminacy issues.}

We start by discussing the properties of network model estimators. Since $P$ is positive-semidefinite by Assumption III, there exists $X$ such that $P = XX^T = \alpha 1_{n}^T + 1_{n} \alpha^T + Z_{12} Z_{12}^T$. Denote $\hX = U_AS_A^{1/2}$ as the ASE of $A$, and $\hZ_{12} = (I- \Pone)\hX$ as the estimated latent factors. We establish the asymptotic distributions for each row of $\hZ_{12}$ in the following theorem.


\begin{theorem}
\label{thm:asymptotic_XZ}
As $n \to \infty$, under Assumptions II-V and further assuming the existence of $\bM_{X} = \lim_{n \to \infty} X^TX/n$ and $ Q_i = \lim_{n \to \infty} \sum_{j \neq i}(x_{i}^T x_{j})(1-x_{i}^T x_{j})x_{j}x_{j}^T/n$ with $x_i$ being the $i$-th row of $X$, for each $1 \leq i \leq n$, we have
$n^{1/2}(\hz_{i, 12} - z_{i, 12}) {\to} \cN(0, \bM^{-1}_{X}Q_i\bM^{-1}_{X})$.
\end{theorem}


Apart from row-wise distributional results, we can also characterize consistency in $2$-$\infty$ norm and Frobenius norm. Specifically, we can show that $\max_{1 \leq i \leq n}\|\hz_{i, 12} - z_{i,12}\|_2 = O_p(1 / \log^{1 + a}(n))$ and $\|\hat Z_{12} - Z_{12}\|_F = O_p(1)$. In \Cref{thm:asymptotic_XZ}, Assumption I is replaced by the assumption on the asymptotic covariance of $X$. Denoting $M_{12} = Z_{12}^TZ_{12}/n$, since we have $M_{X} = M_{12} + \mu_x\mu_x^T$ with $\mu_x = X^T1_{n}/n$, the new assumption is equivalent to assuming that $\lim_{n \to \infty}X^T1_{n}/n$ exists. \Cref{thm:asymptotic_XZ} indicates that the asymptotic variances of $\hz_{i,12}$ and $\hat x_i$ are identical, which corresponds to the intuition that $\hZ_{12} = (I - \Pone)\hat X$ is the centered ASE of $A$.

We also establish the asymptotic distributions for each entry of $\hat \alpha$ in the following theorem.
\begin{theorem}
\label{thm:asymptotic_alpha}
As $n \to \infty$, under Assumptions II-V and further assuming the existence of $\bM_{X} = \lim_{n \to \infty} X^TX/n$ and $\tilde{Q}_i = \lim_{n\to\infty}\sum_{j\neq i}x_{i}^T x_{j}(1-x_{i}^T x_{j})/n$, for each $1 \leq i \leq n$, we have
$ n^{1/2}(\hat{\alpha}_i - \alpha) {\to} \cN(0,  \bM^{-1}_{X} \tilde{Q}_i \bM^{-1}_{X})$.
\end{theorem}


\subsection{Asymptotic Properties of Factor Model Estimators}

By regressing the centered $Y$ on $\hZ_{12}$, we obtain {\color{black} $\hL_{12}^T = (\hZ_{12}^T \hZ_{12})^{-1} \hZ_{12}^T Y$}. 
Owing to the consistency of $\hZ_{12}$, we can establish the asymptotic distribution for each column of $\hL_{12}$ in the following theorem. 
\begin{theorem}
\label{thm:asymptotic_lambda}
As $n \to \infty$, under Assumptions I-V, for each $1 \leq j \leq p$, we have
$n^{1/2}(\hat{\lambda}_{j,12} - \lambda_{j,12}) {\to} \mathcal{N}(0, \bM_{12}^{-1}\Psi_{jj})$,
where $\bM_{12} = \lim_{n \to \infty}  Z_{12}^T Z_{12}/n$ denotes the asymptotic covariance matrix.
\end{theorem}

\Cref{thm:asymptotic_lambda} indicates that $\hl_{j, 12}$ achieves the optimal asymptotic variance because $\bM_{12}^{-1}\Psi_{jj}$ is the variance as if $Z$ is observed in a regression setting rather than being latent. Compared to existing literature, the estimators proposed in \cite{bai2012statistical} can only achieve such optimality under IC*2 or IC*3 in \Cref{tab:ic_bai}, which essentially assume that $Z$ is uncorrelated. On the other hand, our result holds under any set of conditions in \Cref{tab:ic} since $\hZ$ can be estimated from network information. Thus, the generalized factor model requires fewer constraints for optimal inference, which is particularly beneficial for exploratory factor analysis. 

\begin{remark}
The regression with $\hZ$ instead of $Z$ can be viewed as a measurement error problem. We can show that using $\hZ$ for regression induces a finite-sample bias of order $O(n^{-1}\log^{1/2}n)$ for each estimator $\lambda_{j, 12}$, which is negligible for deriving the asymptotic distribution in Theorem \ref{thm:asymptotic_lambda}. More details are provided in the supplementary materials.
\end{remark}

Next, we consider the estimators $\hZ_3$, $\hL_3$, and $\hat \Psi$ obtained from the residual matrix $R$.
We describe their asymptotic results in the following theorem.

\begin{theorem}
\label{thm: estimate_z_3}
Denote $R = (I - \P_{\hZ_{12}} - \Pone)Y$ as the residual matrix and consider the following two cases: (1) if $Z_3$ does not exist, we directly have $\hat\Psi = \Cov{R}$; (2) if $Z_3$ exists, we apply the estimation method described in \Cref{sec:estimate_factor_model} on $R$ to derive $\hZ_3$, $\hL_3$, and $\hat\Psi$. With identifiability conditions IC2* or IC3* in \Cref{tab:ic_bai} for $Z_3, \Lambda_3, \Psi$ and Assumptions I-VI, when $n, p \to \infty$ and $p / n \to c$, where $ c >0$ is a constant, for each $1 \leq i \leq n$ and $1 \leq j \leq p$, we have
\begin{align*}
\sqrt{p}(\hz_{i, 3} - z_{i, 3}) {\to} \mathcal{N}(0, Q^{-1}),
& \quad  n^{1/2}(\hat{\lambda}_{j, 3} - \lambda_{j, 3}) {\to} \mathcal{N}(0, \bM_{3}^{-1}\Psi_{jj}), \\
n^{1/2}(\hat{\Psi}_{jj} - \Psi_{jj}) &{\to} \mathcal{N}(0, 2\Psi^2_{jj}),
\end{align*}
where $Q = \lim_{p \to \infty} \Lambda_3^T \Psi^{-1} \Lambda_3/p$ and $\bM_{3} = \lim_{n \to \infty} Z_3^T Z_3/n$.
\end{theorem}

The asymptotic properties of $\hat\Psi$ remain the same regardless of whether $Z_3$ exists or not. On the other hand, while the convergence rate and asymptotic distribution of $\hL_3$ are similar to $\hL_{12}$ in \Cref{thm:asymptotic_lambda}, the results regarding $\hZ_3$ are different. This difference arises because $\hZ_{12}$ and $\hZ_{3}$ are estimated from distinct models using different estimation methods: $\hZ_{12}$ is estimated from the network using spectral embedding, whereas $\hZ_3$ is estimated from the factor model using the EM algorithm.

\section{Hypothesis Testing of Overlapped Latent Factors}
\label{sec:hypothesis_testing_of_overlapped_latent_structure}


\subsection{Estimating Dimension of $Z_3$}
\label{sec:testing_z_3}
Introducing both individual and shared latent factors presents new challenges for statistical inference in the generalized factor model. In this section, we propose a hypothesis testing framework to address the model selection problems, such as identifying the dimensions of $Z_1$, $Z_2$, and $Z_3$, as well as determining the category each factor belongs to. With the two-step estimation method, we first estimate $k_1+k_2$ from network information. The problem of choosing the dimension of the latent space for the network model has been well-studied in spectral embedding literature \citep{athreya2017statistical}. Thus, we utilize a popular scree-plot-based method   \citep{zhu2006automatic} to estimate $k_1+k_2$ and primarily focus on estimating the dimension of $Z_3$ and differentiating $Z_1$ and $Z_2$ from $\hZ_{12}$ in this work.

We begin by estimating the dimension of $Z_3$. Following the model fitting procedure in \Cref{sec:estimate_factor_model}, this problem is equivalent to estimating the number of factors within $R$. 
We propose a sequential testing procedure that focuses on the gaps among eigenvalues of $\Cov(R)$ to estimate the rank of $\Cov(R)$. Following the high-dimensional factor analysis literature \citep{onatski2009formal}, for each $l \geq 0$, we consider the following testing problem: 
\begin{equation}
    H_0:k_3 = l \quad  \mbox{versus}  \quad H_1: l + 1 \leq k_3 \leq k_{max},
\label{hypothesis:Z_3}
\end{equation}
where $k_3$ denotes the dimension of $Z_3$ and $k_{max}$ is pre-specified as the maximum possible number of factors. We define the test statistic as:
\begin{equation}
    r(l) {=} (\phi_{l + 1} - \phi_{k_{max} + 1}) / (\phi_{k_{max} + 1} - \phi_{k_{max} + 2}),
\label{eq:test_statistics_Z3}
\end{equation}
where $\phi_{l}$ represents the $l$-th largest eigenvalue of $\Cov(R)$. The test statistic $r(l)$ measures the changing ratio of gaps of eigenvalues. Intuitively, a high-dimensional covariance matrix with $l$ latent factors would have the top $l$ eigenvalues diverging to infinity as $(n,p)$ increases, while the remaining eigenvalues stay bounded. Formally, we establish the theorem below.

\begin{theorem}
\label{thm:distribution_test_Z3}
Consider the testing problem \eqref{hypothesis:Z_3} with the test statistic defined in \eqref{eq:test_statistics_Z3}. Under Assumptions I-V, and further assuming $n,p \to \infty$ with $p / n \to c$ where $  c >0$ is a constant, and $\Eps_i$ follows a Gaussian distribution, we have the following results.

(1) Under $H_0$, $r(l) {\to} (x_l - x_{k_{max} + 1}) / (x_{k_{max} + 1} - x_{k_{max} + 2})$, where $x_{l},\ldots, x_{k_{max} + 2}$ jointly follow the Tracy-Widom distribution induced by the Gaussian Orthogonal Ensemble.

(2) Under $H_1$, $r(l) \to \infty$, so asymptotically the testing procedure achieves full power.
\end{theorem}

With \Cref{thm:distribution_test_Z3}, for any $l \geq 0$, we can compare the observed value of $r(l)$
with that simulated from the joint Tracy-Widom distribution \citep{tracy1996orthogonal}.
In practice, we propose setting $k_{max}$ large enough and determining the dimension of $Z_3$ using a sequential testing procedure.
Starting with $l = 0$, 
for each $0 \leq l \leq k_{max}$, if the null hypothesis $H_0: k_3 = l$ is rejected, we further test the null hypothesis $H_0: k_3 = l + 1$ until a null hypothesis is accepted, and then report the corresponding value as the final estimate. {\color{black} The proposed testing procedure can be regarded as a sequence of goodness-of-fit tests for inferring the existence of additional individual factors.  In this way, the dimension of $Z_3$ is estimated through the proposed goodness-of-fit test. In the special case of $l=0$, the proposed test can be used to test the existence of $Z_3$.
}

\subsection{Differentiating $Z_1$ and $Z_2$}

To identify $Z_1$ and $Z_2$ from $\hZ_{12}$, we consider \Cref{eq:linear_regression}, which includes $Z_1$ as part of the regression. If the dimension of $Z_1$ is $k_1$, correspondingly, there would be $k_1$ columns of regression coefficients in $\Lambda_{12}$ with zero vectors. Hence, we can determine $k_1$ by considering the following multiple testing problem. For each $1 \leq l \leq k_1 + k_2$, denoting $\lambda_{\cdot l, 12}$ as the $l$-th column of $\Lambda_{12}$, we consider: 
\begin{equation}
    H_0: \lambda_{\cdot l, 12} = 0_{p \times 1} ~  \mbox{ versus }  ~ H_1: \lambda_{\cdot l, 12} \neq 0_{p \times 1}.
\label{hypothesis:Z_1}
\end{equation}
Under the null hypothesis, the latent factor $l$ belongs to $Z_1$, while under the alternative hypothesis, it belongs to $Z_2$. From \Cref{thm:asymptotic_lambda}, we know that asymptotically each entry $\hat \lambda_{jl, 12}$ follows a normal distribution. Hence, for such a testing problem with a high-dimensional vector, we construct the following sum-of-square test statistics. For each $1 \leq l \leq k_1 + k_2$, define:
\begin{equation}
    S(l) {=}
    \Big(2 {\sum}_{j = 1}^p \hV^2_{jl}\Big)^{-1/2}{\sum}_{j = 1}^p (n \hat \lambda_{jl, 12}^2 - \hV_{jl}),
\label{eq:test_statistics_Z1}
\end{equation}
where $\hV_{jl} = (\hZ_{12}^T\hZ_{12} / n)^{-1}_{ll}\hat \Psi_{jj}$ is a plug-in estimator of the asymptotic variance of $n^{1/2}  \hat \lambda_{jl, 12}$, as specified in \Cref{thm:asymptotic_lambda}. We show the asymptotic distribution for $S(l)$   in the following theorem.

\begin{theorem}
\label{thm:distribution_test_Z1}
Consider the testing problem \eqref{hypothesis:Z_1} and test statistic \eqref{eq:test_statistics_Z1}. Under Assumptions I-V, for each $1 \leq l \leq k_1 + k_2$, we have the following results.

(1) Under $H_0$, $\lim_{n, p \to \infty} S(l) {\to} \cN(0, 1)$.

(2) Under $H_1$, denote the number of non-zero entries in $\lambda_{
\cdot l, 12}$ as $p^{1 - \beta}$. If $\lambda_{jl,12} = \Omega(n^{-1/2}(\log p)^{a})$ holds uniformly with $a > 0$ and $0 < \beta < 1/2$, then we have $S(l) \to \infty$ as $n, p \to \infty$. Hence, asymptotically, the testing procedure achieves full power.
\end{theorem}

{ \color{black} According to \Cref{thm:distribution_test_Z1}, we can test every column of $\hat Z_{12}$ and infer whether it is part of $Z_1$ or $Z_2$, which produces estimations of $Z_1$, $Z_2$ and $k_1$, $k_2$ simultaneously. 
}
In practice, we test multiple latent factors simultaneously. To control the family-wise error rate under multiple testing, we propose using Fisher's method. Specifically, we consider the testing problem:
$H_0: k_1 = 0 ~ \mbox{versus} ~ H_1:k_1 > 0$.
Denote $p(l)$ as the $p$-value derived from the testing problem \eqref{hypothesis:Z_1} for $1 \leq l \leq k_1 + k_2$. Since the $p$-values of these tests are asymptotically independent, with Fisher's method we have that under $H_0$, $- 2 \sum_{l = 1}^{k_1 + k_2} \log p(l) {\to} \chi^2_{2(k_1 + k_2)}$ as $n,p \to \infty$, where $\chi^2_{2(k_1 + k_2)}$ denotes the $\chi^2$ distribution with $2(k_1 + k_2)$ degrees of freedom. While in this work we consider the sum-of-square test statistics that naturally suits the typical dense signal case of factor loading structure, other test statistics proposed under different sparsity levels can also be implemented \citep[e.g.,][]{xu2016adaptive,he2021asymptotically}. {\color{black}The overall estimation and inference procedure is summarized in Algorithm \ref{algo:summary_procedure} below.}

{\color{black}
\begin{algo}
Summary of Estimation and Inference Procedure
\begin{tabbing}
   \qquad \enspace {Step 1: Estimate the Network Model}\\
   \qquad \enspace \quad a. Estimate $\hat k_1 + \hat k_2$ using the scree-plot based method;\\
   \qquad \enspace \quad b. Obtain estimators with spectral embedding:\\
   \qquad \enspace \qquad $\hat Z_{12} = (I - P_1)U_AS_A^{1/2}, ~\text{and} ~\hat{\alpha} = n^{-1}\{I - P_{1}/2\}A1_{n}$;\\
   \qquad \enspace \quad c. Transform the Estimators to Meet Corresponding Identifiability Conditions.\\
   \qquad \enspace {Step 2: Estimate the Factor Model}\\
   \qquad \enspace \quad a. Estimate loadings for $Z_{12}$ with OLS regression: $\hat \Lambda_{12}^T = (\hat Z_{12}^T \hat Z_{12})^{-1} \hat Z_{12}^T Y$;\\
   \qquad \enspace \quad b. Estimate $\hat k_3$ using the testing procedure proposed in Section 4.1;\\
   \qquad \enspace \quad c. Obtain $\hat \Lambda_3$ and $\hat Z_3$ by fitting a factor model on $R = (I - P_{1} - P_{\hat Z_{12}})Y$;\\
   \qquad \enspace \quad d. Separate $\hat Z_1$, $\hat Z_2$ from $\hat Z_{12}$ and obtain $\hat k_1$, $\hat k_2$ with the procedure in Section 4.2.
\end{tabbing}
\label{algo:summary_procedure}
\end{algo}
}


\section{Simulation Studies}
\label{sec:simulation_study}
\subsection{Consistency on Parameter Estimation}


In this section, we use simulation studies to illustrate the finite sample performance of the proposed estimation and testing procedures. We first introduce the simulation settings. 
The network links are generated from the Degree Corrected Stochastic Block Model (DCSBM), a special case of model \eqref{model:net} that accounts for community structure and node degree heterogeneity. Specifically, it models the probability matrix as $P = \rho CWBW^TC$, where $\rho$ is the network density parameter, $C$ is the degree-coefficient diagonal matrix, $W$ is the block assignment matrix, and $B$ is the block-wise link probability matrix. We set all communities to have an equal number of nodes, with $B$ having diagonal block elements equal to $0.8$ and off-diagonal block elements equal to $0.2$. The degree heterogeneity matrix $C$ has its diagonal elements generated as $c_{ii} = 1 / u_i, u_i \overset{i.i.d}{\sim} U(1, 5)$.

After generating $P$, we calculate $Z_{12}$ from $P = \alpha 1_{n}^T + 1_{n} \alpha^T + Z_{12}Z_{12}^T$ with a fixed rotation such that $Z_{12}^TZ_{12}$ is diagonal. For the factor analysis model, entries of $Z_3$ are generated following i.i.d. $\cN(0, 0.2)$, ensuring the variances of $Z_1$, $Z_2$, and $Z_3$ are on the same scale. Entries of the loading matrix $\Lambda$ follow i.i.d. $\kappa \times \cN(0, 1)$, where the multiplier $\kappa$ characterizes the strength of signals, and each diagonal entry of $\Psi$ is generated as $\psi_{jj} \overset{ind.}{\sim} \cN(0, \sigma_j^2)$ with $\sigma_j^2 \overset{i.i.d.}{\sim} U(0.5, 1.5)$. Finally, we transform the generated parameters to meet identifiability conditions $(1)$ described in \Cref{cor:identifiability_of_general_joint_model}. For all experimental settings, we simulate $200$ replicates.


We first illustrate the consistency of the proposed estimators. The reported evaluation criterion is $\mathrm{Tr}(\hat{X}, X) = \mathrm{tr}(X^TP_{\hat{X}} X) / {\mathrm{tr(X^TX)}}$, where $P_{\hat{X}} = \hat{X}(\hat{X}^T\hat{X})^{-1}\hat{X}^T$ is the projection matrix induced by $\hat{X}$. This evaluation criterion can be viewed as a multivariate extension of $R^2$ in linear regression and is widely used as a performance criterion in factor analysis \citep{doz2012quasi}. The optimum of $\mathrm{Tr}(\hat{X}, X)$ is invariant of the dimension of $X$, and a value closer to $1$ indicates a better estimation of $X$. 

\begin{figure}[h]
\centering
    \includegraphics[width = 0.9 \textwidth]{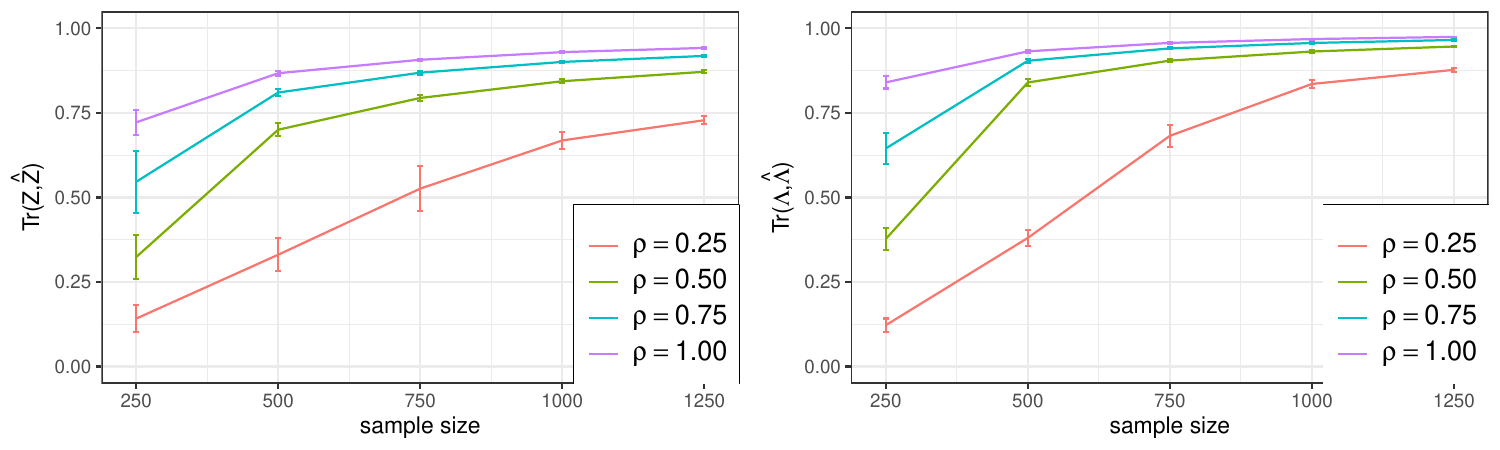}
    \caption{Asymptotic estimation performance on $Z$ (left) and $\Lambda$ (right) with changing network density. Error bars represent one standard deviation.}
    \label{fig:asymptotic_consistency}
\end{figure}

\begin{figure}[h]
\centering
   \includegraphics[width = 0.9 \textwidth]{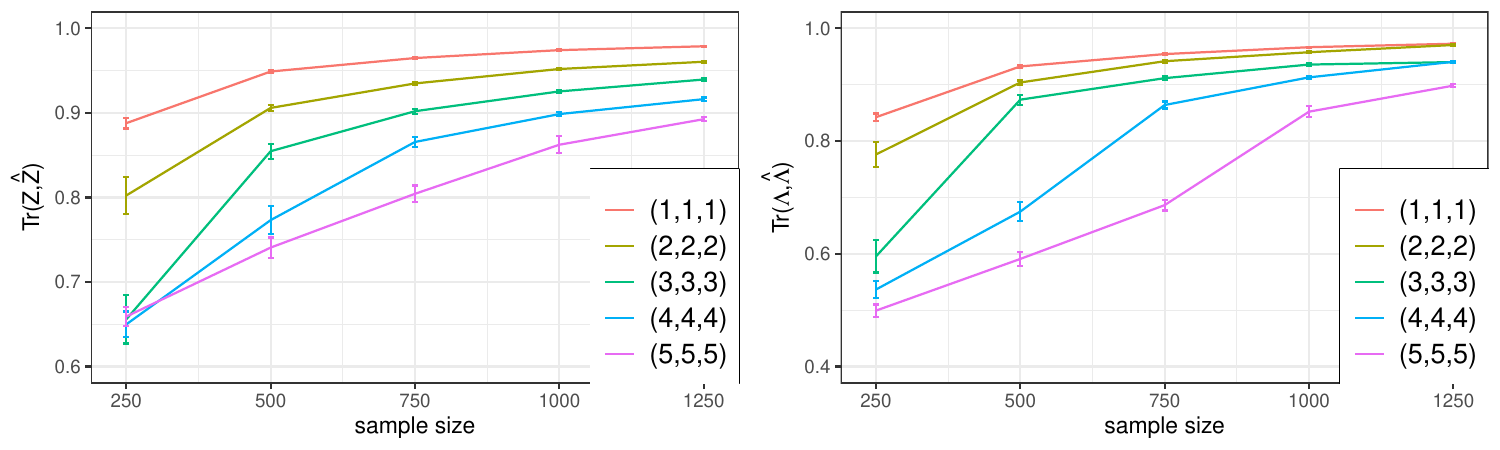}
    \caption{Asymptotic estimation performance on $Z$ (left) and $\Lambda$ (right) with changing number of factors. Error bars represent one standard deviation.}
    \label{fig:consis_many_fact}
\end{figure}

We experiment with column dimensions of $Z_1$, $Z_2$, and $Z_3$ growing from $1, 1$ and $1$ to $5, 5$ and $5$, respectively, and $\kappa = 1, \rho = 1$.  Then we let $n$ and $p$ grow from $250$ to $1250$ with a ratio $p / n = 1$. As shown in \Cref{fig:consis_many_fact}, the estimators become more accurate as the sample size grows. Also, the estimation becomes more challenging with more factors, especially for estimating the network model with a smaller number of samples. We also study how network density affects estimation accuracy. We fix the number of factors to be $1, 3$ and $1$ for $Z_1$, $Z_2$, and $Z_3$, respectively, $\kappa = 1$, $p/n = 1$, while letting $\rho$ vary from $0.25$ to $1$. When $\rho = 1$, the empirical network density is around $0.05$. \Cref{fig:asymptotic_consistency} illustrates that the estimators are more accurate with a larger network density parameter, as more information becomes available in this case.

In the supplementary material, we provide additional results on the effect of network density when changing the ratio between $p$ and $n$. The results indicate that the effect of network density varies little with different $p/n$ ratios. We also include extra experiments comparing the accuracy of $\hZ_{12}$ and 
$\hL_2$ estimated from binary data $A$ versus $\hZ_3$ and $\hL_3$ estimated from continuous variables $Y$, corresponding to Theorems 2, 4 and 5.

\subsection{Power and Accuracy of Testing Procedures}
\label{sec:power_and_accuracy_of_testing_procedures}
To study the power of the proposed testing procedures, we fix the dimensions of $Z_1, Z_2$ and $Z_3$ to be $1, 3$ and $1$, respectively, with $n = p = 500$, $\rho = 1$, while allowing the signal strength of $\Lambda$, $\kappa$, to vary from $0$ to $0.3$. We set the test significance level to be $0.05$.  Since no existing methods in the literature consider overlapped latent factors, we compare the proposed method with the oracle case where $\hZ$ is replaced with true values of $Z$.

\begin{figure}[ht]
    \centering
    \includegraphics[width = 0.9 \textwidth]{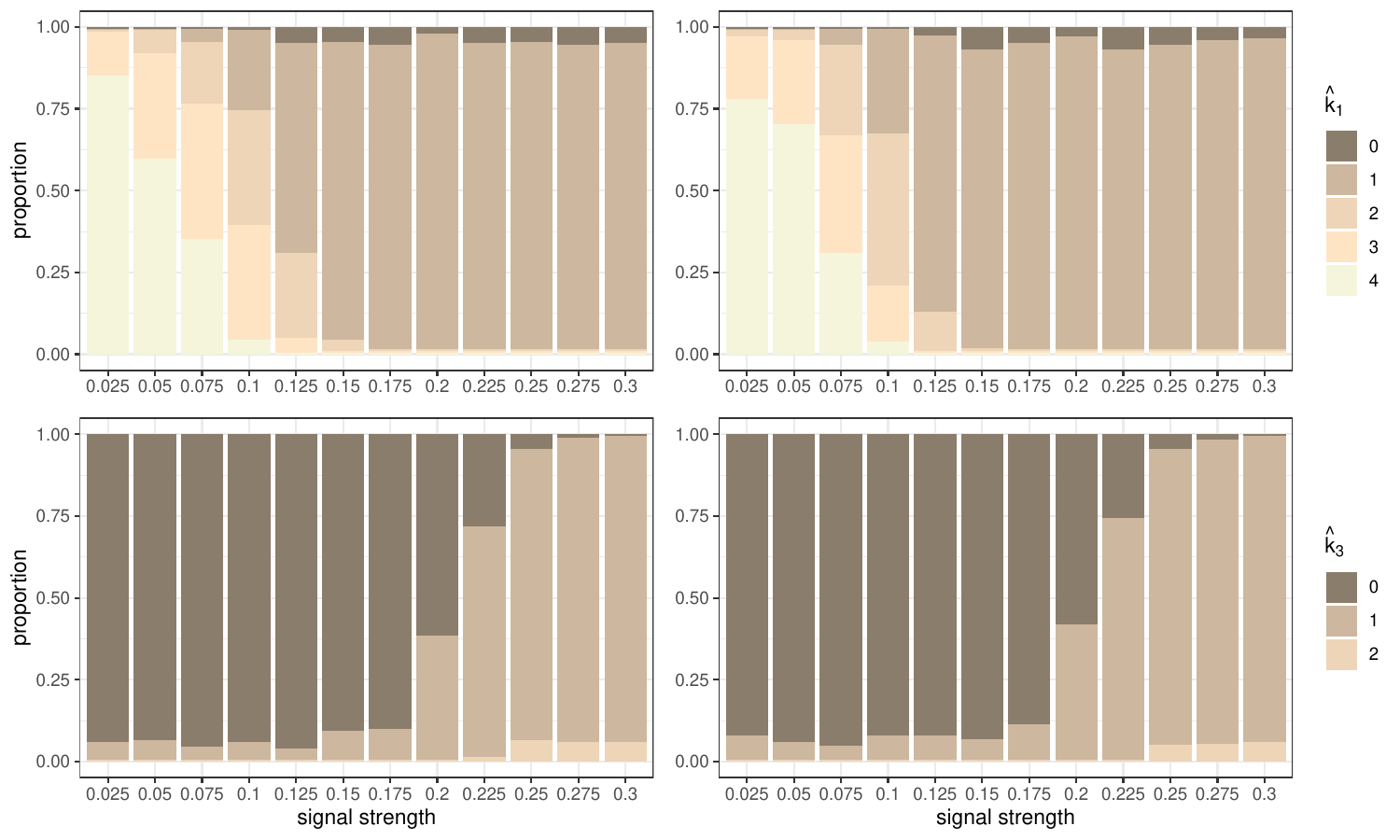}
    \caption{Estimation of $k_1$ (top) and $k_3$ (bottom), with test statistics calculated from estimated latent factors (left) versus true values (right). Both $k_1, k_3$ have true values of $1$.}
    \label{fig:estimated_dimension}
\end{figure}

As illustrated in \Cref{fig:estimated_dimension}, when signal strength $\kappa$ is small, the testing procedure fails to identify the factor model structure of $Z_2$ and $Z_3$ in most cases.
As signals become stronger, our method correctly recovers the truth of $k_1 = 1, k_3 = 1$, and the performance is comparable to the oracle tests. Note that the average estimated dimensions would deviate from the truth by a small fraction even under strong signal cases. Such deviation is the consequence of making type-I errors and can be theoretically quantified as the level of significance.

We further investigate the accuracy of the proposed testing procedure on estimating the number of factors. To estimate $k_3$, we follow the sequential procedure in \Cref{sec:testing_z_3}. When estimating $k_1$, we find that directly reporting the total number of rejected individual tests would underestimate $k_1$ for two reasons: (1) the finite-sample bias of $\hZ_{12}$ and $\hL_{12}$ becomes larger as estimating a network model with more latent factors becomes more difficult; and (2) there are finite-sample correlations among the test statistics. Thus, in practice, we propose to implement the following permutation method: first, randomly permute the rows of $Y$ and calculate the test statistics for a large number of times ($500$ in our case), and then derive empirical quantiles of the test statistics.

\begin{table}[h!]
\centering
\begin{tabular}{|ccc|ccc|}
\hline
$(k_1, k_2,k_3)$ & $\hat k_1$  & $\hat k_3$  & $(k_1, k_2,k_3)$ & $\hat k_1$  & $\hat k_3$  \\ \hline
(1, 1, 1) & 0.96 (0.20) & 1.05 (0.22) & (3, 1, 1) & 2.83 (0.74) & 1.08 (0.27) \\
(1, 1, 3) & 0.94 (0.25) & 3.04 (0.18) & (3, 1, 3) & 2.89 (0.31) & 3.07 (0.26) \\
(1, 3, 1) & 0.93 (0.26) & 1.05 (0.22) & (3, 3, 1) & 2.85 (0.36) & 1.05 (0.22) \\
(1, 3, 3) & 0.92 (0.27) & 3.04 (0.20) & (3, 3, 3) & \multicolumn{1}{c}{\cellcolor[HTML]{F8F8F8}{\color[HTML]{333333} 2.86 (0.60)}} & 3.04 (0.20) \\
\hline
\end{tabular}
\caption{Estimation of $k_1$ and $k_3$ with changing number of factors, both means and standard deviations are reported.}
\label{tab:test_multi_factors}
\end{table}

We set $n = p = 500, \rho = 1, \kappa = 1$, while changing $k_1, k_2$, and $k_3$ between $1$ and $3$. 
The means and standard deviations of estimated dimensions for $Z_1$ and $Z_3$ are reported in \Cref{tab:test_multi_factors}. We can see that both the sequential test for $Z_3$ and the permutation test for $Z_1$ work well under all combinations of $k_1, k_2$ and $k_3$. Specifically, when $k_1 = 3$ the average number of $\hat k_1$ is around $2.85$, which is the expected value of $\hat k_1$ under the significance level $0.05$, provided that the three test statistics are independent. {\color{black}More experiments on estimation and testing procedures can be found in the supplementary material.}

\section{Analysis on Statistician Coauthorship Data}
\label{sec:real_data_analysis}

{\color{black}

\begin{figure}[t]
  \centering
  \includegraphics[width=0.475 \linewidth]{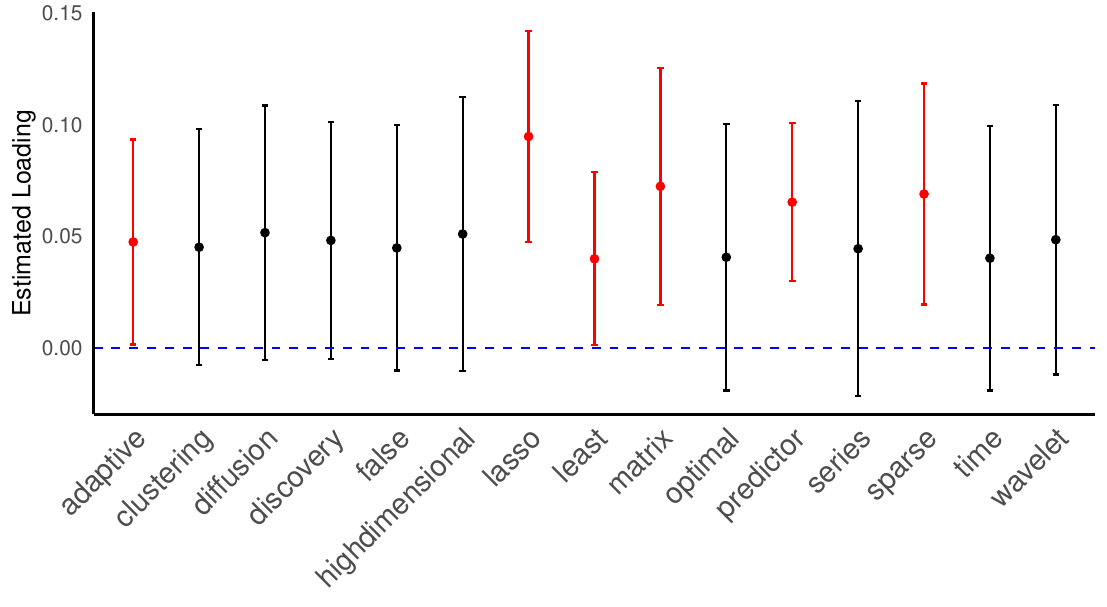}
  \hfill
  \includegraphics[width=0.475 \linewidth]{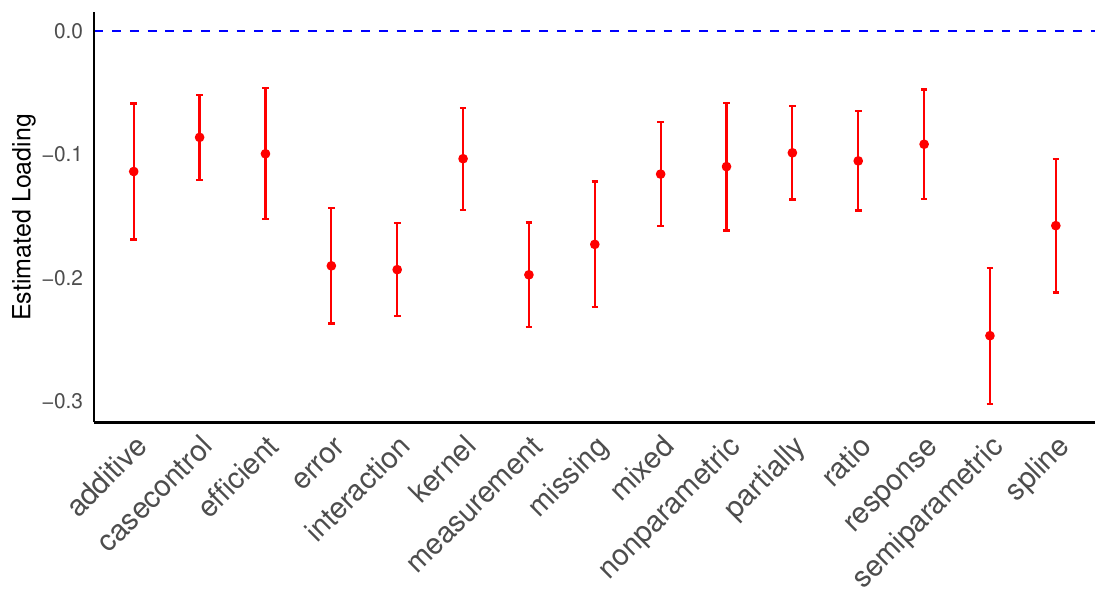}
  \caption{Top 15 terms associated with factor 1, on the positive (left) and negative (right) loadings.}
  \label{fig:topic_terms}
\end{figure}

With the generalized factor model, we analyze the statistician coauthorship network dataset collected in \cite{ji2016coauthorship}, which contains titles and authors of publications from four popular statistical journals during 2003-2012. After preprocessing, the dataset contains $n=1011$ nodes, where two statisticians are connected if they have collaborated on one or more recorded publications. Each node is associated with $p=283$ covariates that reflect the importance (i.e. adjusted usage frequency) of statistical terms.  Our goal is to explore potential latent factors associated with each statistician's preference on statistical terms and collaboration patterns among them. Details of the preprocessing procedure can be found in the supplementary material.
} 


We first explore the dataset by listing the statistical terms with the top importance scores. Some of them are closely related to specific research topics: Bayesian, clustering, functional, high-dimensional, longitudinal, semiparametric, Markov, nonparametric, selection, survival, etc. Many others are general statistical terms: asymptotic, confidence, consistency, estimator, likelihood, optimal, regression, etc. The coauthorship network is plotted in the supplementary material, where we can see some noticeable hub nodes (statisticians with many collaborations). 

{\color{black} We investigate the association between the two types of data by fitting the proposed model. Through the testing procedure, we identify $7$ factors for the network data, which are all shared with node covariates ($Z_2$) with p-values less than 1e-8 and none is individual ($Z_1$). We also find no individual factors for node covariates ($Z_3$) with a p-value of 0.726. These findings indicate a strong association between the frequency of statistical terms used by statisticians and their collaboration patterns.}
We further study the association by interpreting the shared latent factors and the corresponding loading matrix. The estimated latent factors can be naturally interpreted as statistical research topics, which statisticians would have varying degrees of interest in (characterized via $Z$) and statistical terms would have different levels of relevance to (characterized via $\Lambda$). We follow the factor analysis literature \citep{harman1976modern} to rotate latent factors and loading matrix with $varmax$ method, which introduces sparse structure in the loading matrix for more interpretable results. 

{\color{black}
\Cref{fig:topic_terms} illustrates top 15 terms that have the largest loadings associated with factor 1 on both the positive and negative sides. A $95\%$ confidence interval is constructed for every term, where terms having significantly nonzero loadings are coloured in red. From \Cref{fig:topic_terms} we can see that most of the significant loadings are on the negative part, of which the top terms are closely related to the research topic ``Nonparametric / Semiparametric Statistics". In the supplementary material, we present the interpretation for the rest of factors, as well as the names of the statisticians with top magnitudes of weights on the same factors, which in combination can provide more meaningful interpretations of the statistical communities in different research areas. We further conduct a missing value imputation experiment to demonstrate that incorporating network information helps predict node covaraites, whose results can also be found in the supplementary material.
}

\section{Discussion}
\label{sec:discussion}
This work presents a generalized factor model designed to accommodate high-dimensional variables and network links among observations simultaneously. Through theoretical analysis and a real data example, we demonstrate that incorporating network information into the high-dimensional factor model is advantageous both in theory and practice. Notably, the proposed estimation method follows a two-step approach. Alternatively, one could jointly estimate all parameters by maximizing the likelihood of $A$ and $Y$ simultaneously. Although the joint estimation approach may potentially yield more accurate estimations of shared latent factors, it is less flexible in incorporating additional data. In cases where network information is updated over time or additional individual variables are collected, the two-step estimation approach only requires refitting part of the model. 

{\color{black} Similar to RDPG, the network model in Equation \eqref{model:net} assumes positive definiteness of the probability matrix $P$, which limits its model expressiveness. One potential improvement is to incorporate the Generalized RDPG \citep{rubin2017statistical} network model that allows $P$ to be an indefinite matrix. Allowing negative eigenvalues of $P$ would make the model more flexible while also more challenging to interpret, and we leave it for future exploration.

In the factor analysis literature, there exist works that consider the shared and individual factor structures among multiple tabular datasets, such as \cite{andreou2019inference} and \cite{lock2013joint}. However, both of these two works consider data of same tabular formats and that every dataset follows the same factor model in Equation (1). Contrastingly, in this work we consider network data with node covariates, which are of different data formats and thus require different model forms. Such divergence introduces unique challenges and leads to differences in model assumptions and methodologies. Specifically, in the context of identifiability conditions, the authors of \cite{lock2013joint} adopted the identifiability conditions that $Z_2$ being orthogonal to $(Z_1, Z_3)$. Given that we are more interested in the benefits of introducing network information for factor model, we find it more natural to assume $Z_{12}$ to be orthogonal to $Z_3$. Furthermore, regarding the methodology in inferring the shared and individual factors, \cite{andreou2019inference} considered the problem with datasets of uniform formats with same factor models and proposed a CCA-based method that can only determine the number of shared factors. In comparison, our proposed procedure developed for network data with node covariates is capable of not only identifying the number of shared factors but also specifying which columns represent these shared factors.
} 

The proposed joint modeling framework, featuring shared and individual latent factors, can be naturally extended to model more complex data. For example, instead of modeling network links as a binary adjacency matrix, we can propose a latent space model for general relational data \citep{hoff2018additive} encoded as a weighted adjacency matrix. Moreover, we can include not only continuous but also categorical high-dimensional variables. 
While this work primarily focuses on high-dimensional variables and considers network links as auxiliary information, the joint modeling framework is not limited to such settings. By shifting the focus, high-dimensional variables may help address challenges in statistical inference on large-scale and sparse network data.

\bibliographystyle{biometrika}
\bibliography{paper-ref}

\end{document}